\documentstyle[11pt,ihepconf,twoside,epsfig]{article}
\begin{document}
\newcommand{\be}{\begin{equation}}
\newcommand{\ee}{\end{equation}}
%\thispagestyle{empty}
%\phantom{.}
%\large
\vskip 1cm
\begin{center}
{\Large{\bf ENERGY IN  THEORY OF GRAVITY \\AND ESSENCE OF TIME \\}}
\vspace*{1cm} {\bf Ivanhoe B. Pestov}\\
{\it  Bogoliubov  Laboratory of Theoretical Physics, Joint Institute
for Nuclear Research,\\  Dubna,  Russia }\end{center}
\vskip 2cm

{\small
In the framework of the field theory it is shown that a time
(viewed as a scalar temporal field) is an internal property of the physical
system, which defines its causal structure and evolution.
A new concept of internal time allows to solve the energy problem in
General Relativity and
predicts the existence of matter outside the time.
It is demonstrated that introduction of the
temporal field permits to derive the physical laws of the electromagnetic
field( the general covariant four dimensional Maxwell equations for the
electric and magnetic fields) from the geometrical equations of this field.
It means that the fundamental physical laws are in full correspondence with
the essence of time. On this ground, from the geometrical laws of the
gravitational field the physical evolution equations of this
field are derived. Two characteristic solutions of these equations are
obtained (including the Schwarzschild solution).   
}
\medskip

\bigskip

\section{Introduction}
In the theory of gravitational field the problems
connected with  the energy conservation exist in a literal sense since
the time of its creation when
Einstein set up the problem of including gravity into the framework of the
Faraday concept of field. Thorough and deep
analysis of the problem of gravity field given by him in the works [1, 2]
enables to formulate in what follows the key principles of gravity physics
(General Relativity). However, till now in the framework of these principles 
there is no adequate solution to the energy conservation problem [3, 4]. 
This topic is still the subject of active research and as this implies, 
still only incompletely understood. This means that the theory of gravity 
field is not unique. The fact that the theory is not unique is a bad feature, because
if theory is not unique it is clearly missing some essential elements.
 
In the presented paper we find these elements and give a simple solution of
the problem in question,
which is based on the connection between the time and energy and
necessarily follows from the first principles of General Relativity if one
puts them into definite logical sequence. The energy conservation means
that the rate of change with time of total energy density of gravity field
and all other fields is equal to zero and hence energy density represents
the first integral of the system.

New understanding of time presented here will have
implications in quantum gravity. In fact,
a major conceptual problem in this field is the notion of time  and
how it should be treated. The importance of this issue
was recognized at the beginning of the history of quantum gravity [5], but
the problem is still unresolved and has drawn recently an increasing
attention (see, for example~[6], [7] and [8] and further references therein).

The paper is organized as follows. In \S2 we formulate
the fundamentals of general covariant theory of time with the guiding
idea that the  manifold is the main notion in physics and that time itself
is a scalar field on the manifold which defines the evolution of the full
system of fields being one of them. The system of interacting fields is
considered to be full if the gravitational field is included in it.

All known dynamical laws of nature have the following form:
the rate of change with time of certain quantity  equals to the results of
action of some operator on this quantity. So, a general covariant
definition of rate of change  with time of any field
is one of the main results of the theory of time presented here. It is
also a starting point and relevant condition for the consideration of the
concept of evolution and of the problem how to write the field equations in
the general covariant evolution form.

In \S3 the connection of the temporal field with
the Einstein gravitational potential is established. This gives
the possibility to derive the geometrical laws of the gravitational
field, general covariant law of energy conservation and produce the general
covariant expression for the energy density of the gravity field. The whole
approach relies on the concepts of genuine Riemann geometry and the physical
and geometrical sense of Lorentz signature is recognized.
To demonstrate a concrete application of
general theory, in \S4 it is shown how evolution equations for
the vectors of electric and magnetic fields in the four dimensional
general covariant form can be derived from the geometrical equations for the
bivector of electromagnetic field.
Through this it is shown that the original Maxwell
equations (which as a matter of fact express the fundamental physical law)
are in full correspondence with fundamentals of the theory of time.
In \S5 the notion of the momentum of the gravitational field is
introduced and dynamical equations of this field is deduced from the
geometrical laws established in \S3. In \S6 the exact solutions
of these equations are considered. One of them describes the gravitational
field of pointlike gravitational charge and other one represents the
field of the homogenous and isotropic distribution of the gravitational
energy. Some concluding remarks are given in \S7.

\section{Central role of time in gravity theory}

In accordance with the principles of gravity physics, in this section
the fundamentals of the theory of time are formulated.

According to Einstein, in presence of gravitational field all the systems of
coordinates  are on equal footing and in general, coordinates have
neither physical nor geometrical meaning. Thus, in the gravity
theory the coordinates play the same role as the Gauss coordinates in his
internal geometry of surfaces from which all buildings of the modern
geometry are grown.
Hence, one needs to construct the internal theory of physical fields
analogous to the Gauss internal geometry of surfaces. In other words, the
problem is how the modern differential geometry transforms into the
physical geometry.

The fields characterize the events and fill in the geometrical space.
In view of what has been said above, this space is a four dimensional
smooth manifold because this structure does not distinguish intrinsically
between different coordinate systems (the principle of general covariance
is naturally included into this notion).  Hence it follows that the notion
of smooth manifold is the primary issue not only in differential geometry
but also in the theoretical physics ( dealing with gravitational
phenomena.  This means that all other definitions, notions and laws should
be introduced into the theory through the notion of smooth manifold. Indeed
if some notion, definition or law is in agreement with the structure of
smooth manifold, then they are general covariant, i.e., do not depend on
the choice of coordinate system.

Definition of manifold is considered to be known and we only notice
that  all information on this topic can be found for example in reference
[3] or [9]. For our purposes it is enough to keep in mind that all smooth
manifolds can be realized as surfaces in the Euclidean space.
In what follows we shall consider only four dimensional manifolds. That is
evident from the physical point of view. However there is also a deep purely
mathematical reason for this choice.
Smooth manifold consists of topological manifold and
differential structure defined on it.
It is known [9], that a topological manifold always admits differential
structure if and only if its dimension is not larger than four.

It is clear that, in general, manifold should be in some relation
with its material content i.e., the fields. 
In view of this it is very important to know how
material content designs manifold. If we take the point of view that
manifold apriori is arena for the physical events, then it
is natural to select simplest manifold, for example, manifold of special
theory of relativity. However, it is evident that nature of gravity field
is not compatible with  an apriori defined manifold.

It can be shown that manifold as a surface in the Euclidean space is
designed by the covariant symmetrical tensor field  $g_{ij}$  on the
manifold, for which  adjoined quadratic differential form (Riemann metric)
\begin{equation}
ds^2 = g_{ij} du^i du^j
\end{equation}
is positive definite.
Thus, covariant positive definite symmetrical tensor field
$g_{ij}(u)$ is the necessary element of any  general covariant
intrinsically self-consistent physical theory. Here we only give the
defining system of differential equations $$g_{ij}(u^1, \, u^2, \, u^3,
         \, u^4) = \delta_{ab} \frac{\partial F^a}{\partial u^i}
\frac{\partial F^b}{\partial u^j}, \quad a,b = 1, \cdots ,4+k, \, k \geq 0
  $$ avoiding any detailed consideration.  If the functions
$g_{ij}(u^1,\, u^2, \, u^3, \, u^4)$ are known in local system of
coordinates $u^1, u^2, u^3, u^4$, then solving this system of equations we
obtain the functions $F^a(u^1, \, u^2, \, u^3, \, u^4)$ and hence the
region of manifold, defined by the equations $x^a = F^a(u^1, \, u^2, \,
u^3, \, u^4), $ where $x^a $ are the Cartesian coordinates of embedding
Euclidean space.

An important conclusion that follows from this consideration is that
there is one and only one way for the other fields to design manifold which
can be explained as follows. Let $u^1, u^2, u^3, u^4$  be a local
system of coordinates in the vicinity of some point of an abstract
manifold. Let us determine in such a vicinity the system of
differential equations that connect basis field $g_{ij}(u)$ (genuine 
Riemann metric) with other
ones. Solving this system of equations we find field $g_{ij}(u)$ and by
doing so, we design a local manifold of physical system in question. In what
follows, a smooth manifold that corresponds to a physical system will be
called a physical manifold.

For further consideration of the principles of internal field theory
we simply note that there is a fundamental difference
between physics and geometry. In geometry there is no motion
that is tightly connected with the concept of time.
Thus, to be logical, we need to introduce time into the theory using
its first principles. Since, in general,
coordinates have no physical sense, time can be presented as a set of
functions of four independent variables (or in more strict manner as a
geometrical object on the manifold). It is quite obvious from the logical
point of view.

We put forward the idea that the time is a scalar field on
the manifold. By this we get a simple answer to the
question with long standing history "What is time ?" It should be
emphasized, that temporal field (together with other fields) designs
manifold as it was explained above but it has also another functions which
will be considered below.

Temporal field with respect to the coordinate system
$ u^1, u^2, u^3, u^4 $ in the region $U$ of smooth four dimensional manifold
$M$  is denoted as $f(u) = f(u^1, u^2, u^3, u^4).$ If the temporal field
is known, then to any two points $p$ and $q$ of manifold one can put in
correspondence an interval of time \begin{equation} t_{pq} = f(q) - f(p) =
\int \limits_{p}^{q} \partial_i f du^i.  \end{equation}

Unlike time, space is not an independent entity.
Instead of space we shall consider space cross-sections of the  manifold
$f^{-1} (t),$ which are defined by the temporal field.  For
the real number $t,$  space cross-section is defined by the equation
\begin{equation}
f(u^1, u^2, u^3, u^4) = t. \label{3}
\end{equation}
One can call the number $t$ "the height" of the space cross-section
of manifold. If a point $p$ belongs to the space cross-section $f^{-1}
(t_1),$ and a point $q$  to the space cross-section $f^{-1} (t_2),$ then
the time interval (2) is equal to the difference of the heights  $t_{pq} =
t_2 - t_1.$ It is clear that $t_{pq}=0 $ if $p$ and $q$ belong to the same
space cross-section. Thus, to the every  segment of curve one can put in
correspondence a time interval.

Given the general covariant definition of time, one should show
that it is constructive in all respects. First of all we consider how the
temporal field defines the form of physical laws. It is known that the
general form of physical laws is very simple and is based on the following
recipe:  the rate of change  with time of a certain quantity is
equal to the result of action of some operator on this quantity. To be
concrete, let us consider Maxwell equations. We know that the rate of
change with time of electrical and magnetic fields enter the dynamical
equations of electromagnetic field. Thus, we need to give a general
covariant definition of the rate of change with time of any field and in
particular this definition should be applicable for the case of
electromagnetic field.

This problem has fundamental meaning, since it is impossible to speak
about physics when one has no mathematically rigorous definition of
evolution. It is clear that correct general covariant definition should
be conjugated with simple condition: if the rate of change with time of
some field is equal to zero in one coordinate system, then in any
other coordinate system the result will be the same.

On the manifold there is only one general covariant operation that can be
considered as a basis for the definition of the rate of change with time
of any field quantity. This general covariant operation is called
derivative in given direction and is defined by the vector field on
the manifold and the structure of the manifold itself. Thus, the problem is
to connect the temporal field  $f(u^1, u^2, u^3, u^4)$ with some vector
field $t^i (u^1, u^2, u^3, u^4).$  Since a temporal field is a scalar
one, the partial derivatives  define covector field $t_i = \partial_i
f.$  Now, the following definition becomes self-evident:
the gradient of temporal field (or the stream of time) is the
vector field of the type 
\be
t^i = (\nabla f)^i = g^{ij} \frac{\partial
f}{\partial u^j} = g^{ij} \partial_j f = g^{ij} t_j, \label{4}
\ee
where $g^{ij}$ are the contravariant components of the Riemann metric (1).
The gradient of the field of time defines the direction of the most rapid
increase (decrease) of the field of time. We define now the rate of change
with time of some quantity  as the derivative in the direction of
the gradient of the field of time and denote this operation by the symbol
$D_t$.

Let us find the expression for the rate of change with time of the
temporal field itself. We have,
$$D_t f =t^i \partial_i f = g^{ij}
\partial_i f \partial_j f. $$
Since  $ D_t f$ is a general covariant generalization of the evident
relation $\frac{d}{dt} \, t = 1,$  the temporal field should obey the
fundamental equation
 \be
(\nabla f)^2 = g^{ij} \frac{\partial f}{\partial u^j}
\frac{\partial f}{\partial u^j} =1. \label{5}
\ee
Equation (5) means that the rate of change with time of the temporal field
is a constant quantity, the most important constant of the theory. From the
geometrical point of view the equation (5) simply shows that the gradient
of the temporal field is unit vector field on the manifold with respect
to the scalar product that is defined as usual by the metric (1), $ (V, W)
= g_{ij} V^i W^j .$

It should be noted that the equation $(\nabla f)^2 = 1 $ is the main
equation of the geometrical optics. In view of this one can consider
equation (5) as the equation of 4-optics. This analogy can be
useful for consideration of some special problems in the theory of
time.

The rate of change with time of the  symmetrical tensor field
 is  given by the expression
\begin{equation} D_tg_{ij} =  t^k \frac{\partial
 g_{ij}}{\partial u^k} + g_{kj} \frac{\partial t^k}{\partial u^i} +
 g_{ik} \frac{\partial t^k}{\partial u^j} \ .   \label{6}
 \end{equation}
Similar formulas can be presented for any other geometrical quantities.
In mathematical literature the derivative with respect to the given
direction is usually called the Lie derivative.  Thus, one can say that the
rate of change  with time of any field  is the Lie derivative with respect
to the direction of the stream of time.

Consider the notion of time reversal and the invariance with respect to this
symmetry that is very important for what follows. It is almost evident that
in general covariant form  the time reversal invariance   means that theory
is invariant with respect to the transformations
\be t^i \rightarrow  -  t^i. \label{7} \ee
It is clear that theory will be time reversal invariant if the
gradient of temporal fields will appear in all formulae only as an even
number of times, like $t^i t^j.$

Within the scope of special relativity and quantum mechanics, time and
energy are tightly connected. It is natural to  suppose that in gravity
physics the link between time and energy  even more deep and energy
conservation follows from the invariance of the Lagrangian theory with
respect to the transformations \be f(u) \rightarrow f(u)+a , \ee where $a$
 is arbitrary constant.  This invariance means that all the space sections
of manifold of system in question are equivalent.

Einstein himself put in correspondence to the gravity field symmetrical
tensor field ${\tilde g}_{ij},$ which is characterized by the condition
that adjoined quadratic differential form
\begin{equation}
d{\tilde s}^2 ={\tilde g}_{ij} du^i du^j,   \label{9}
\end{equation}
has the signature of the interval in special relativity. In accordance with
the principle of gravity physics discussed above, it is natural to
assume that Einstein's interval (9) has a structure that is defined by
the form-generating field $g_{ij}$ (Riemann's metric (1)) and temporal
field. If disclosed, this structure will give a simple method to introduce
temporal field into the equations of gravitational physics. It is quite
evident that the metric (1) has an Euclidean signature and hence it has no
structure by definition. To be transparent in our consideration, let us
remind a simple mathematical construction.  One can consider tensor field 
$S^i_j$ as linear transformation ${\bar V}^i = S^i_j V^j $ in the vector 
space in question.
If the operator $S$ is selfadjoint, that is $(V, SW) = (SV, W),$ then it is
always possible to introduce the scalar product associated with operator
$S$ via the formula $<V,V> = (V, SV).$ It is clear that  associated scalar
product will be in general indefinite and with respect to the initial
scalar product it has a structure. Thus, in general, the connection between
the forms (1) and (9) is given by the relation $ {\tilde
g}_{ij} = g_{ik}S^k_j.  $ We shall give now simple expression
for the operator  $S^i_j,$  which defines the Einstein interval and is a
simple generalization
of time reversal invariance
(7) as a fundamental physical principle.  To this end, we shall define
$T$- symmetry in the space of the vector field in order to
have transformation (7) as a particular case. We say that vector
fields ${\tilde V}^i$ and $V^i $ are $T$- symmetrical, if the sum of
this fields  is orthogonal to the gradient of temporal field and their
difference is collinear to it,
$ ({\tilde V}^i+ V^i) t_i =
0, \quad {\tilde V}^i - V^i = \lambda t^i.  $
We have,
${\tilde V}^i = V^i - 2 n^i (V ,  n) = (\delta^i_j - 2n^i n_j) V^j,$
where
$$ n^i = \frac{t^i}{\sqrt{(t,t)}}, \quad (t,t) = g_{ij} t^i t^j.$$
From this formula it follows that the fields  $ t^i$ and $ -t^i$ are
$T$-symmetrical and hence the definition of $T$-symmetry given in the
space of the vector field is correct.
From the above consideration it follows that  $S$ is operator of
$T$- symmetry  that is $S^i_j = \delta^i_j - 2n^i n_j $ and hence
for the Einstein's potential  we obtain the following expression
\be {\tilde g}_{ij} = g_{ik} (\delta^k_j - 2n^k n_j) = g_{ij} - 2 n_i
n_j, \label{10} \ee which is invariant under the transformations (7).  The
contravariant components of the tensor field $\tilde g_{ij}$ are
 $\tilde g^{ij} = g^{ij} - 2 n^i n^j,$  $ \tilde g^{ik} \tilde g_{jk} =
\delta^i_j.$ 

Let us give the physical meaning of the Einstein's scalar product associated
with $T$- symmetry. Since
$$ < V, V>= (V,V) - 2 (V,n)^2 =
|V|^2 (1- 2 \cos ^2 \phi) = - |V|^2 \cos2\phi ,$$
where $\phi$  is the
angle between the vectors $V^i$ and $n^i$, the Einstein's scalar
product is indefinite and can be positive, negative or equal to zero
according to the value of the angle $\phi.$ In particular,
$ <V, V > =0,$ if $\phi = \pi/4.$ Thus, the Einstein's
scalar product is time reversal invariant and permits to classify all the
vectors depending on which  angle they form with the gradient of the
temporal field.  As we see, the temporal field and T-symmetry define the
Einstein's form (9) as the metric of the normal hyperbolic type.  Hence,
the gradient of the temporal field defines the causal structure on the
physical manifold and can be identified with it. It is the physical meaning
of the Einstein's interval and mechanism of formation of physical
pseudo-Euclidian metric as well. It should be noted that in our reasoninges
we did not use the equation (5). This will become clear later under the
consideration of the variational principle.

Now we shall show that the causal structure (the gradient of the temporal
field) can be reduced to the canonical form $(0,0,0,1)$ by the suitable
coordinate transformation at once in all points of some (may be small)
patch of any point on the physical manifold.  Local coordinates with
respect to which gradient of the temporal field has the form $(0,0,0,1)$
will be called compatible with causal structure or intrinsic coordinates.
It should be noted very important significance of the system of coordinates
compatible with causal structure since it is similar to the Darboux system
of coordinate in the theory of symplectic manifolds which is a geometrical
basis for the Hamilton mechanics. What is more, there are transformations
of coordinates which conserve the causal structure and these
transformations are analogous  to the canonical transformations in
Hamiltonian mechanics.

Geometrically the stream  of time is defined as a congruence of lines (lines
of time) on the manifold.  
Analytically the lines of time are defined as the solutions of the
autonomous system of differential equations
 \begin{equation} \frac{du^i}{dt} = \; g^{ij}
\frac{\partial f}{\partial u^j} = g^{ij} \partial_j f = (\nabla f)^i ,
\quad (i=1,2,3,4). \label{11} \end{equation}
Let
\be u^i(t) = \varphi^i(u^1_0,\, u^2_0, \, u^3_0, \, u^4_0, t) =
\varphi^{i}(u_0,t) \label{12} \ee
be the solution to equations (11) with initial data
$\varphi^{i}(u_0, t_0) = u^i_0 $ so that
$$ {\partial {\varphi^i(u_0, t_0)}}/~{\partial u^j_0} = \delta^i_j.  $$
Substituting  $u^i(t) = \varphi^i(u_0, t)$ into the function
$f(u^1, \,u^2,\,u^3,\, u^4)$ we obtain
$p(t) = f(\varphi(u_0,t)).$ Differentiating this function with respect to
$t,$ by virtue of (5) and (11), one finds
$dp(t)/dt = 1.$ It leads to
$ f(\varphi(u_0,t)) = t - t_0 + f(u_0). $
Suppose that all initial data belong to the space section
$ f(u^1_0,\, u^2_0, \, u^3_0, \, u^4_0 ) = t_0.  $
Rewriting this relation in the parametric form
$u^i_0 = \psi^i(x^1, x^{2}, x^3) ,$ Eqs.
(12) can be written as the system of relations
 \be u^i = \phi^i (x^1, x^{2}, x^3, t). \label{13}\ee
The functions (13) have continuous partial derivatives with
respect to variable  $x^1, x^{2}, x^3, t $ and their functional determinant
is not equal to zero. Hence the functions $ \phi^i (x^1, x^{2}, x^3, t)$
design intrinsic  system of coordinate of the dynamical system in
question.  Now one can show that in such a system of coordinates the
covariant and contravariant components of the gradient of the temporal
field and some components of the field $g_{ij}$ take a simple numerical
value $$ t^i =(0,0,0,1) = t_i, \quad g_{44} = g^{44} =1, \quad g_{\mu4} =
g^{\mu4} =0, (\mu =1,2,3)$$ and hence this coordinate system is compatible
with causal structure.  
Therefore, it
follows that in the system of coordinates compatible with causal structure,
metric  (1) takes the form\be ds^2 = g_{\mu\nu}(x^1, x^2, x^3, t ) dx^{\mu}
dx^{\nu} + (dt)^2, \quad \mu,\nu =1,2, 3 ,\ee since $t_i = g_{ij} t^j =
g_{i4}.$

So, for any point of physical manifold we can indicate a local
coordinate neighborhood with the coordinates compatible with causal
structure. Covering manifold with such charts we get atlas on the
manifold that is compatible with its causal structure. In this atlas the
field equations have the most simple form in the  sense that
all components of the gradient of temporal field   take numerical values.
We see that causal structure is analogous to the symplectic
structure  of the Hamiltonian mechanics.

From the above consideration it also follows that variable $t$,
parametrizing the line of time, can be considered as
coordinate of time of the physical system in question. This name is
justified particularly by the fact that
the rate of change with time of any field  is equal to the partial
derivative with respect to $t$, i.e., $D_t = \partial / \partial t $ in the
system of coordinate compatible with causal structure (see for example
(6)).  This is very important point since we see that the intrinsic
coordinates give the possibility to write all equations in the
form analogous to the canonical form of the Hamilton equations.

Furthermore, if we reverse time putting
${\tilde t}^i = - t^i,$ then the lines of time will be parametrized by
new variable  ${\tilde t.}$ From the equations (11) it is not
difficult to derive that there is one-to-one and
mutually continuous correspondence between the parameters  $t$ and ${\tilde
t}$  given by the relation
${\tilde t} = -t.$ From here it is clear that in the system of
coordinate defined by the time reversal, the variable $-t$ will be the
coordinate of time. Thus, the general covariant definition of time reversal
given by (7) is adjusted with the familiar definition that is connected with
the transformation of coordinates. Finally, we note one more important
fact related to the intrinsic coordinates. Transformations~(8) take the
well-known form $t \rightarrow t + a $ in the system of coordinates
compatible with the causal structure.

\section{Geometrical laws of the gravitational field}

Consideration made above gives the evident geometrical method of
constructing Lagrangians in gravity physics. It is clear that geometrical
laws of the gravitational field are defined by the Lagrangian $ L = {\tilde
R} ,$ where ${\tilde R} $ is a scalar that is constructed from the
Einstein's gravitational potential in the standard way.

Since the Einstein interval is defined by the two fields connected by the
equation (5), it is necessary to pay special attention when deriving the
equations of gravitation field. A standard method is to incorporate the
constraint (5) via a Lagrange multiplier $\varepsilon = \varepsilon(u)$,
rewrite the action density  for gravity field in the form  $ L_g = {\tilde
R} + \varepsilon(g^{ij} t_i t_j - 1)$ and treat the components of the
fields $g_{ij}$ and $ f$  as independent variables.

Let us consider the action
\be A= \frac{1}{2} \int {\tilde R} \sqrt g \,d^{4}u + \int  L_m(\tilde g, F)
\sqrt g \,d^{4}u + \frac{1}{2} \int \varepsilon(g^{ij} t_i t_j - 1) \sqrt g
\, d^{4}u, \label{15} \ee
where $g = Det(g_{ij}) > 0$ and $L_m(\tilde g, F) $
is the Lagrangian density of the system of other fields $F$
which incorporates the Einstein's gravitational potential in the
conventional form. Such geometrical method of introduction of causal
structure into the geometrical equations of  fields does not
require special explanation but it is not evident how from geometrical laws
to derive general covariant dynamical laws ( the form of which was defined
earlier).  This
problem will be solved here in the two important cases. It  will be
demonstrated how from the geometrical laws of the electromagnetic field to
derive the Maxwell equations for the electric and magnetic fields in
general covariant four-dimensional form (notice that this problem remains
unresolved up-to-date).  In other case we shall derive general covariant
dynamical laws of the gravitational field from the geometrical laws of this
field.  

Since Einstein's gravitational potential is the function of
$g_{ij}$ and  $f$, it is necessary to use the chain rule. 
We have, $\delta \tilde R = {\tilde g}^{ij} \delta {\tilde R}_{ij} +
{\tilde R}_{ij} \delta {\tilde g}^{ij}. $ Further we denote the
Christoffel symbols as
 $ \Gamma^i_{jk}$ and $ {\tilde \Gamma}^i_{jk}$
belonging to the fields $g_{ij}$ and
$ {\tilde g}_{ij}$, respectively. In what follows, the covariant derivatives
with respect to $ \Gamma^i_{jk}$ ( $ {\tilde \Gamma}^i_{jk}$) will be denoted as
$\nabla_j $ (${\tilde \nabla}_j $).
One can show that ${\tilde
g}^{ij} \delta {\tilde R}_{ij}  $ can be omitted as a perfect differential.
Varying now ${\tilde g}^{ij}, $ we get $$\delta {\tilde g}^{ij} = \delta
g^{ij} + P^{ij}_{kl}\delta g^{kl} + Q^{ijk} \partial_k \delta f ,$$ where
$$P^{ij}_{kl} = 2 n^i n^j n_k n_l - n^i (n_k \delta^j_l + n_l \delta^j_k) -
n^j (n_k \delta^i_l + n_l \delta^i_k), $$
$$Q^{ijk} =\frac{2}{\sqrt {(t,t)}}(2n^i n^j n^k  - n^i g^{jk} -n^j g^{ik}
) \ .$$
A tensor field  $P^{ij}_{kl}$  is symmetrical in covariant and
contravariant indices.  Let $ G_{ij} = {\tilde R}_{ij}- \frac{1}{2} {\tilde
g}_{ij} {\tilde R}$ be the Einstein's tensor and further we put
$\delta L_F = \frac{1}{2} M_{ij} \delta {\tilde g}^{ij}, $ and
introduce by  the standard way the energy-momentum tensor ${T}_{ij} =
M_{ij} - {\tilde g}_{ij}L_F.  $ 
Observing that $${\tilde
g}_{ij} + {\tilde g}_{kl}P^{kl}_{ij} = g_{ij}, \quad g_{ij} Q^{ijk} = n_i
n_j Q^{ijk}=0, $$ it is easy to verify that a  total variation of the action
can be presented in the following form
\be \delta A = \frac{1}{2} \int (A_{ij})\delta g^{ij} +
B \delta f +(g^{ij} t_i t_j -1) \delta \varepsilon) {\sqrt g} d^{4}u ,
\label{16} \ee where $$A_{ij} = G_{ij} + G_{kl} P^{kl}_{ij}+ {T}_{ij}
+{T}_{kl} P^{kl}_{ij} + \varepsilon t_i t_j - \varepsilon (g^{kl} t_k t_l
-1) g_{ij}),$$ $$B = - \nabla_k( (G_{ij} + T_{ij} - \varepsilon  t_i t_j )
Q^{ijk}) - 2 \nabla_k ( \varepsilon t^k) \ . $$
One can consider tensor $P^{kl}_{ij} $
as operator  $P$ acting in the space of symmetrical tensor fields. The
characteristic equation of this operator has the form $P^2 + 2P = 0 ,$ and
hence $(P+1)^2 =1.$ Thus, operator $P+1$ is inverse to itself. Since  $t_i
t_j + t_k t_l P^{kl}_{ij} = -t_i t_j , $  from (16) it follows that in
framework of considered here concept of time  the geometrical Einstein 
equations have the form \be G_{ij} + {T}_{ij} = \varepsilon {\partial_i}f
{\partial_j}f , \quad g^{ij} {\partial_i}f {\partial_j}f =1 \label{17},\ee
\be \nabla_k (\varepsilon t^k ) = 0 \label{18}. \ee
The equations (17)  constitute the full system of geometrical equations
of the gravitational field.  As it is shown above, these equations emerge
from the first principles of gravity physics.

Equation (18) expresses the law of energy conservation in gravitational
physics which, evidently, is general covariant.
To make sure that we indeed deal with conservation of energy, it is
sufficient to figure out that action (15) is invariant with
respect to transformation (8) and hence the equation~(18) results also from
the Noether's theorem. It is also clear that the Lagrange multiplier 
$\varepsilon $ has a physical meaning of energy density of the system 
in question. From
the Eqs. (17) it follows that \be \varepsilon = G_{ij}t^i t^j + {T}_{ij}
t^i t^j  = \varepsilon_g + \varepsilon_m ,\label{19} \ee where $\varepsilon_g =
G_{ij}t^i t^j = \frac{1}{2}{\tilde R}_{ij} g^{ij}$ is energy density of the
gravitational field and  $ \varepsilon_m = {T}_{ij} t^i t^j $ is  energy
density of other fields.  Consider the law of energy conservation from the
various points of view. First of all we consider link between (17) and
(18). The so-called local energy conservation is written as follows  $
{\tilde \nabla}_i {T}^{ij} = 0,$ where ${T}^{ij} = {T}_{kl} {\tilde g}^{ik}
{\tilde g}^{jl}.$ These equations are fulfilled on the equations of the
fields $F$ that contribute to the energy-momentum tensor.  Since $ {\tilde
\nabla}_i G^{ij} = 0$ identically, from 
(17) it follows that $$ {\tilde \nabla}_i {T}^{ij} = \varepsilon t^i
({\tilde \nabla}_i t^{j}) + t^j {\tilde \nabla}_i (\varepsilon t^i) .$$ 
Since ${\tilde \nabla}_i
(\varepsilon t^i)= \nabla_i (\varepsilon t^i) $ and 
${\tilde \nabla}_i t^{j}=0, $ finally we have $$
{\tilde \nabla}_i { T}^{ij} = t^j \nabla_i (\varepsilon t^i) .$$  In view
of this the energy conservation law can be treated as the condition of
compatibility of the field equations. In this sense, the  law of energy
conservation is  analogous to the law of charge conservation.

Show that rate of change with time of the energy density
$D_t (\sqrt g \varepsilon) $  equal to zero
 $D_t (\sqrt g \varepsilon) =0$ and hence this quantity is a first integral
 of the system in question.  We have $D_t (\sqrt g \varepsilon) = t^i
 \partial_i (\sqrt g \varepsilon) + \sqrt g \varepsilon \partial_i  t^i =
\partial_i (\sqrt g \varepsilon t^i).$  Since $  {\sqrt g} \nabla_k
(\varepsilon t^k ) = \partial_i (\sqrt g \varepsilon t^i),$  from the
law of energy conservation (18) it follows that energy density is the first
integral of the system $$ D_t (\sqrt g \varepsilon) = 0.  $$ In the system
of coordinates, compatible with causal structure, this equation has a more
customary form $$ \frac{\partial}{\partial t} (\sqrt g \varepsilon) = 0. $$

The Eqs. (17)  are geometrical and from the physical point of view it is
not evident that they express the fundamental dynamical laws of the nature.
This is not a simple exercise to derive from~(17) a physical laws of the
gravitational field and we first consider more easy (but very
important as well) problem.  We derive the general covariant Maxwell
equations from some natural geometrical equations.

 \section{Time in the theory of electric and magnetic fields}

Let $A_i $ be the vector potential of the electromagnetic field.
We define the gauge invariant tensor of electromagnetic field as usual
$ F_{ij} = \partial_i A_j - \partial_j A_i$ and according to the
principle of gravity physics introduce temporal field into the
theory of electromagnetic field through the gauge invariant and
general covariant Lagrangian $$
L_{em} = \frac{1}{4} F_{ij} F_{kl} {\tilde g}^{ik}  {\tilde g}^{jl}. $$
with the energy-momentum tensor
\be T_{ij} = F_{ik} F_{jl} {\tilde g}^{kl}
- {\tilde g}_{ij} L_{em} \label{20} \ .\ee
The equations for the
tensor $F_{ij}$ can be written in the form
\be  \nabla_i {\stackrel{*} F}{^{ij}} = 0 , \quad \nabla_i {\tilde F}^{ij} =
0, \label{21} \ee where $$ {\stackrel{*} F}{^{ij}} = \frac{1}{2} e^{ijkl}
F_{kl}, \quad {\tilde F}^{ij}= F_{kl} {\tilde g}^{ik} {\tilde g}^{jl}, $$
and $e^{ijkl} $ are contravariant components of the Levi-Civita tensor
normalized as $e_{1234} = \sqrt g.$ Now we shall derive from geometrical
laws of the electromagnetic field the general covariant  four dimensional
Maxwell equations for the electric and magnetic fields which in fact
express the fundamental physical laws.  First of all we formulate the main
relations of the vector algebra and vector analysis on the four dimensional
physical manifold which is interesting by itself.

A scalar product of two vector fields
$A^i$ and $B^i$ is defined by the Riemann metric (1) as earlier.
A vector product of two vector fields  $A^i$ and $B^i$ we shall
construct as follows   $$ C^i = [A B]^i
=e^{ijkl}t_j A_k B_l \ .  $$ Here the crucial significance of the
temporal field should be emphasized. It is evident that $[A B] + [B A] = 0.$
By the direct
calculation it can be shown that $|[A B]| = |A| |B| \sin \varphi, \quad
[A[B C]] = B (A,C) - C(A,B).$

Differential operators of the vector analysis on the physical manifold are
defined as natural as algebraic ones. For the divergence and gradient we
have respectively   $$  div \,A =   \nabla_i A^i  = \frac{1}{\sqrt g}
\partial_i (\sqrt g A^i),\quad( grad \,\phi )^i = ( g^{ij}-t^i t^j)
\partial_j \phi = g^{ij} \partial_j \phi - t^i D_t \phi  $$ and as a
consequence  $$ div \, grad \phi = \frac{1}{\sqrt g}
\partial_i (\sqrt g g^{ij} \partial_j \phi) - \frac{1}{\sqrt g} \partial_i
(\sqrt g t^{i} D_t \phi)= \nabla_i \nabla^i \phi - \nabla_i (t^i D_t \phi).
$$
A rotor of the vector field $A$ is defined as a vector product of $ \partial
$ and $A$ $$ ( rot\,A )^i = e^{ijkl}t_j
\partial_k A_l = \frac{1}{2} e^{ijkl}t_j ( \partial_k A_l -
\partial_l A_k ).$$ It is easy to verify that $$
rot \, grad \, \phi = 0.$$
It is evident that all operations so defined are general covariant.

There is only one  direct way to derive from the geometrical Eqs. (21) the
fundamental physical laws  first formulated by Maxwell. Consider
the rate of change with time  of the potential of electromagnetic field.
By definition, we have $$ D_t A_i = t^k \partial_k A_i + A_k \partial_i t^k
= t^k (\partial_k A_i - \partial_i A_k) + \partial_i(t^k A_k) = t^k F_{ki}
- \partial_i \phi,$$  where $\phi = - t^k A_k. $ Thus, the rate of change
with time of the electromagnetic potential can be presented as the
difference of two covector fields with one of them having the form \be E_i
= t^k F_{ik} \label{22}. \ee From this it follows that strength of the
electric field is general covariant and gauge invariant quantity that is
defined by the equation $$ E_i =t^k F_{ik} = - D_t A_i - \partial_i \phi.
$$ Now it is quite clear that general covariant and gauge invariant
definition of the magnetic field strength is given by the formula $$ H^i =
(rot A)^i = e^{ijkl}t_j \partial_k A_l = \frac{1}{2}
e^{ijkl}t_j ( \partial_k A_l - \partial_l A_k ) .$$ Thus, \be H_i
= t^k {\stackrel{*} F}_{ik} \label{24},\ee where ${\stackrel{*} F}_{ij}
= g_{ik} g_{jl} {\stackrel{*} F}{^{kl}}. $

Now it is not difficult to derive the Maxwell equations for the electric
and magnetic fields
from the geometrical Eqs.  (21).
Resolving equations (22), (23) over  $F_{ik},$ we obtain \be
F_{ij} = - t_i E_j + t_j E_i - \varepsilon_{ijkl} t^k H^l .\label{24}\ee
Thus, on the physical manifold there is general covariant one-to-one
algebraic relation between the electric and magnetic fields and tensor of
the electromagnetic field that is defined by the temporal field.
Out of  (24) we find \be {\stackrel{*} F}{^{ij}} = - t^i H^j + t^j H^i -
e^{ijkl} t_k E_l , \quad   {\tilde F}^{ij} =  t^i E^j - t^j E^i -
e^{ijkl} t_k H_l .   \label{25} \ee
Substituting (25) into (21), we shall obtain the
Maxwell equations for the strengths of the electric and magnetic
fields in the following general covariant and gauge invariant form
\be \frac{1}{\sqrt g}  D_t(\sqrt g  H)  = -
rot E , \quad   \frac{1}{\sqrt g} D_t(\sqrt g  E)  = rot H . \label{26}\ee
From Eqs. (26) it follows that $$ D_t (\partial_i(\sqrt
g H^i)) = 0 , \quad D_t (\partial_i(\sqrt g E^i)) = 0 , $$ and
therefore the dynamical equations are compatible with constraints.

To complete this discussion with the Maxwell equations as the
main topic, we write the expression for the components
of the energy-momentum tensor in terms of electric and magnetic field
strength \be
T_{ij}= \frac{1}{2} g_{ij} (|E|^2 + |H|^2) - E_i E_j - H_i H_j -t_i {\Pi}_j
-t_j {\Pi}_i,  \label{27}\ee where ${\Pi}_i$ are covariant components of the
Pointing vector $$
{\Pi}^i= e^{ijkl} t_j E_k H_l,\quad { \Pi}=[{ E} { H}].$$
From (27) and definition given above
we find the energy density of the electromagnetic field
 $$ \varepsilon_{em} = \frac{1}{2} (E^2 + H^2) $$
which is a reasonable result. We shall formulate also the
energy conservation law starting from the general covariant Maxwell
equations.
Out of (26) and (27) we get
\be
\frac{1}{{\sqrt g}} D_t ({\sqrt g} \varepsilon_{em} ) + \nabla_i {\Pi}^i = -
\frac{1}{2} T^{ij} D_t g_{ij}, \label{28} \ee
where $T^{ij} = {\tilde g}^{ik} {\tilde
g}^{jl} T_{kl}.$ Let us now assume that the electromagnetic field
is considered on the background of the physical manifold of some
full system of fields. Let it be known also that for gravity field of
this system the equation $D_t g_{ij} =0$  holds valid.  In such
approximation, when physical manifold is external with respect to the
electromagnetic field, from (28) we obtain that the energy density of the
electromagnetic field satisfies the equation  $$ \frac{1}{{\sqrt g}} D_t
({\sqrt g} \varepsilon_{m} ) + \nabla_i {\Pi}^i = 0. $$ It is exactly the
energy conservation law of the electromagnetic field in the above mentioned
approximation.

Thus, it is shown that the principles of the theory of time and gravity
physics are in full correspondence with the fundamental physical laws and
hence they can be considered as a method to derive  new fundamental
equations.

\section{Physical laws  of the gravitational field}
Now we have an important experience that allows us to derive 
physical equations of the gravitational field
analogous to the Maxwell equations. The guiding idea is very simple. The
equation $div \,E =4\pi \rho$ is tightly connected with the charge
conservation.  Thus, our goal is to find gravitational analog of this
equation which should be connected with the energy conservation in a manner
similar to the conservation of the electric charge.

To this end we first of all introduce the important notion
of the momentum of gravitational field, i.e. quantity which is analogous to
$E_i = D_t A_i + \partial_i (t^i A_i).$  Consider a tensor
field of the type~(1,1)
$$P^i_j = \frac{1}{2} g^{ik} D_t g_{jk} = \frac{1}{2} g^{ik} (\nabla_j t_k
+ \nabla_k t_j ) = g^{ik} \nabla_j t_k =  \nabla_j t^i.$$
A tensor field so defined will be called the momentum of the gravitational
field or simply the momentum of the field. Give some important
relations for the momentum $$   t_i P^i_j =0, \quad  t^j P^i_j
=0, \quad  g^{ik} P^l_k g_{lj}= P^i_j .  $$ $$   D_t P^i_j =
 t^l \nabla_l P^i_j,  \quad \nabla_i P^k_j- \nabla_j P^k_i =
R_{ijl}{^k} t^l. $$ $$  D_t {\Gamma}^{i}_{jk} = \nabla_j P^i_k
+ \nabla_k P^i_j - g^{il} g_{km}\nabla_l P^m_j, \quad D_t g =  g
g^{ik} D_t g_{ik} = 2 g P^i_i .  $$ As the following step in
the required direction  we shall write geometrical equations (17) in other 
form.

By contraction with ${\tilde g}^{ij}$  we get from (17) that
 ${\tilde R} = \varepsilon + T, $  where $T = T_{ij} {\tilde g}^{ij}. $
Using this relation we transform (17) to the following form \be  {\tilde
R}_{ij} + T_{ij}- \frac{1}{2} {\tilde g}_{ij} T = \frac{1}{2} g_{ij} \,
\varepsilon \label{29} .  \ee If we start with equations (29), we get by
contraction (19) and ${\tilde R} = \varepsilon + T, $   and so we can get
back to (17).  We may  use either (17) or (29) as the basic equations.

With (5)  we get
${\tilde R}_{ij} = R_{ij} +  \nabla_l (t^l D_t g_{ij})$
and hence
$g^{ik} {\tilde R}_{jk} = R^i_j +  2 D_t P^i_j + 2 P^k_k P^i_j .$
Introduce the tensor fields
$$B^i_j = h^i_k R^k_l h^l_j +   D_t P^i_j +  P^k_k P^i_j,\quad
 N^i_j =  h^i_k T^k_l h^l_j - T h^i_j,$$
where $h^i_j = \delta^i_j - t^i t_j $ is projection operator, $h^i_k h^k_j
= h^i_j.$   We next define the flux vector of the energy of the
gravitational field $$G_i  =\nabla_k P^k_i - \partial_i P^k_k + t_i
(P^k_l P^l_k + D_t P^k_k )$$  and the flux vector of the energy of other
fields $${\Pi}_i =  \varepsilon_m \, t_i - T_{ik}t^k. $$ It is evident that
$(t,G)= (t, \Pi) = 0.$  With this we can derive from the equations
(29) the following system of equations
\be D_t P^i_j + P^k_k P^i_j + B^i_j
+ N^i_j = \frac{1}{2} \varepsilon \, h^i_j  \label{30} \ee
\be  G_i  = {\Pi}_i  \label{31} \ee
and vice versa. Our statement is that equations (30) represent in general
covariant form the physical laws of the gravitational field similar to
the Maxwell equations (26). To emphasize this analogy we write equations
(30) in the form
$$ \frac{1}{\sqrt g}  D_t(\sqrt g  P^i_j ) + B^i_j
+ N^i_j = \frac{1}{2} \varepsilon  \, h^i_j  $$
and add that like equation $ div E = 0,$  the equation (31) may be
considered as constraint since from (30) it can be derived that
$$D_t (G_i - {\Pi}_i) + P^k_k (G_i - {\Pi}_i) =   \frac{1}{\sqrt g}
D_t(\sqrt g (G_i - {\Pi}_i)) = 0 \ .  $$  The last equation means also that
equations (30) and (31)  are compatible.

Equation (31) has very simple physical sense
that  the  flux vector of the energy of the gravitational field is exactly
equal to the flux vector of energy of other fields .
For the energy density of the gravitational field    we have $
{\varepsilon}_g = \frac{1}{2}{\tilde R}_{ij} g^{ij} = \frac{1}{2} R +
(P^i_i)^2 + D_t P^i_i = T + U,$ where \be T = \frac{1}{2} (P^i_i)^2 -
\frac{1}{2} P^i_j P^j_i \label{32} \ee is a density of kinetic energy and
\be {U} = \frac{1}{2} R + \frac{1}{2} (P^i_i)^2 +  \frac{1}{2} P^i_j P^j_i
+ D_t P^i_i \label{33} \ee is a density of potential energy  of the
gravitational field.

For the better understanding of the last constructions we shall
consider the general covariant dynamical laws of the gravitational
field in the atlas compatible with causal structure. With respect to the
coordinate transformations that conserve causal structure the quantity
$g_{\mu\nu}(t,x^1,x^2,x^3), \quad (\mu, \nu = 1,2,3)$ from the metric (14)
transforms as  the symmetrical tensor and $dl^2 =
g_{\mu\nu}(t,x^1,x^2,x^3) dx^{\mu} dx^{\nu} $ may be considered as the
metrics of the space sections of the physical manifold. For this one
parametric set of three dimensional metrics we can consider Christoffel
symbols $L^{\mu}_{\nu\sigma}$ and covariant derivative $\nabla_{\mu}$,
Riemannian tensor curvature $S_{\mu\nu\sigma}{^{\tau}},$ tensor Ricci
$S_{\nu\sigma} = S_{\mu\nu\sigma}{^{\mu}} $  and scalar curvature $ S =
g^{\mu\nu} S_{\mu\nu} = S^{\mu}_{\mu} $  as usual. In the atlas compatible
with causal structure we have $L^{\mu}_{\nu\sigma} = {\Gamma}^{\mu}_{\nu
\sigma}, $ $P^4_{i} = P^i_4 =0,$ $B^4_{i} = B^i_4 =0, \, (i=1,2,3,4) $ and
what is more important that $$ B^{\mu}_{\nu} =S^{\mu}_{\nu},  \quad R +
(P^i_i)^2 +  P^i_j P^j_i + 2D_t P^i_i  = S  \ .$$ Thus, out of (32) and (33) we
get for the densities of the kinetic  and potential energy of the
gravitational field following representation $$ T =
\frac{1}{2}(P^{\sigma}_{\sigma})^2 - \frac{1}{2} P^{\mu}_{\nu}
P^{\nu}_{\mu}, \quad  U = \frac{1}{2} S.$$

Now we write equations (30) and (31) in the atlas
compatible with causal structure under assumption that $T_{ij}$ is the
tensor of the energy--momentum of the electromagnetic field. In this case
$$N^{\mu}_{\nu} = T^{\mu}_{\nu} = \frac{1}{2}(E^2 + H^2)
\delta^{\mu}_{\nu} - E^{\mu} E_{\nu} - H^{\mu} H_{\nu}$$ is the Maxwell
stress energy tensor. Thus, we can write the basic equations (30) and (31)
in the following form \be {\dot P}^{\mu}_{\nu} + P^{\sigma}_{\sigma}
P^{\mu}_{\nu} + S^{\mu}_{\nu} + T^{\mu}_{\nu} = \frac{1}{2} \varepsilon \,
{\delta}^{\mu}_{\nu} , \label{34} \ee \be  \nabla_{\nu}
P^{\nu}_{\mu} - \partial_{\mu} P^{\sigma}_{\sigma} = {\Pi_{\mu}} =[{\bf E}
{\bf H}]_{\mu} \label{35},\ee where dot stands for partial derivative
with respect the variable $t.$  From equations (34) and (35) it follows that
the Cauchy problem for the gravitational field is not more difficult than
for the electromagnetic field.

In the above consideration we derived the law of energy conservation from
very general geometrical laws.  It is very important from the physical
point of view to consider energy conservation on the basis of Maxwell
equations and physical laws of gravitational field (34) and (35).  In the
atlas compatible with causal structure the Maxwell equations (26) read
$${\dot E}^{\mu} + P^{\sigma}_{\sigma} E^{\mu} = e^{\mu\nu\sigma}
\partial_{\nu} H_{\sigma}, \quad {\dot H}^{\mu} + P^{\sigma}_{\sigma}
H^{\mu} =-- e^{\mu\nu\sigma} \partial_{\nu} E_{\sigma}.$$ Since ${\dot g} =
g g^{\mu\nu}  g_{\mu\nu} = 2 g P^{\sigma}_{\sigma},$  $
\frac{\partial}{\partial t} (\sqrt g \, \varepsilon ) =\sqrt g ({\dot
\varepsilon} + P^{\sigma}_{\sigma} \varepsilon  ) $
and  one can derive from these
equations the formula $${\dot \varepsilon_{em}} + P^{\sigma}_{\sigma}
\varepsilon_{em}  = - \nabla_{\nu} {\Pi}^{\nu} - T^{\mu}_{\nu}
P^{\nu}_{\mu}.  $$

In the atlas compatible with causal structure we have for the energy
density of the gravitational field  $${\dot {\varepsilon}_g} =
P^{\sigma}_{\sigma} {\dot P}^{\sigma}_{\sigma}- P^{\nu}_{\mu} {\dot
P}^{\mu}_{\nu} + \frac{1}{2} {\dot S}.$$ From the equations (34)
we have the relations $${\dot P}^{\sigma}_{\sigma} =
\frac{1}{2} \varepsilon_{em} - \frac{1}{2} \varepsilon_{g}  - P^{\nu}_{\mu}
P^{\mu}_{\nu}, \quad P^{\nu}_{\mu} {\dot P}^{\mu}_{\nu} = \frac{1}{2}
{\varepsilon} P^{\sigma}_{\sigma}- P^{\sigma}_{\sigma} P^{\nu}_{\mu}
P^{\mu}_{\nu} - P^{\nu}_{\mu} S^{\mu}_{\nu} - P^{\nu}_{\mu} T^{\mu}_{\nu}.
$$ With this one can show that $${\dot \varepsilon}_g + P^{\sigma}_{\sigma}
\varepsilon_g  = \frac{1}{2} g^{\mu\nu} {\dot S}_{\mu\nu} + T^{\mu}_{\nu}
P^{\nu}_{\mu}. $$ Since $\frac{1}{2} g^{\mu\nu} {\dot S}_{\mu\nu}  =
\nabla_{\nu} {G}^{\nu}, $  for $\varepsilon = \varepsilon_{g}
+\varepsilon_{em}$  we have $${\dot \varepsilon} + P^{\sigma}_{\sigma}
\varepsilon = \nabla_{\nu} {G}^{\nu}  - \nabla_{\nu} {\Pi}^{\nu} .  $$
Thus, \be \frac{\partial}{\partial t} ({\sqrt g} \varepsilon ) =
\partial_{\nu}({\sqrt g} G^{\nu} ) -  \partial_{\nu}({\sqrt g}
\Pi^{\nu}) \label{36}.\ee  We may integrate this relation over a
three-dimensional volume $V$ lying in the space section of the physical
manifold. For the energy in the volume $V$ we have   $W = \int \varepsilon
{\sqrt g} dV.  $ Consider the  fluxes of the gravitational $\Phi_g = \int
\Omega_{\nu} d{\sigma}^{\nu} $ and the electromagnetic $\Phi_{em} = \int
\Pi_{\nu} d{\sigma}^{\nu} $ energy, where integral is taken over the
boundary surface of the  volume $V.$  Since the right hand side of the
equation (36) can be converted by Gauss's theorem, from (36) we have
the relation $$ \frac{\partial}{\partial t} W  = \Phi_g - \Phi_{em} . $$
Thus, the law of energy conservation means that the fluxes of the
gravitational and electromagnetic energy flow in opposite directions and
exactly equal to each other.  ( We know that our sun is a surprisingly
powerful source of the radiant energy.  Now it is evident that if there is
the flow of gravitational energy in  the direction to the sun then we can
say that our sun is the transducer  of the gravitational energy into the
electromagnetic one). Now we consider question about exact solutions of
the equations (34) and (35).

\section{Exact solutions}
We have seen above that there is deep analogy between the energy and charge
and the gravitational and electromagnetic fields. Here we consider the
existence of solutions to the equations (34) and (35) which
reproduce the physical situation with pointlike distribution
of the gravitational energy, i.e.
$\varepsilon_g = m \delta(r),$ where $m$ is a gravitational charge of
point source.  This physical situation is analogous to the pointlike
distribution of density of charge, when $\rho = q \delta(r),$ where $q$ is
a electric charge of point source. Thus, we search the solution of the
gravitational equations analogous to the Coulomb potential.  Starting point
is the question about static gravitational field.

Gravitational fields will be called {\bf
static} if the rate of change  with time of the gravitational
potential is equal to zero, $D_t g_{ij} = 0.$ In accordance with (6), this
condition can be written as $$ D_t g_{ij} = \nabla_i t_j + \nabla_j t_i
=0.  $$ It is evident that definition of the static gravitational fields is
general covariant.  Since $\nabla_i t_j - \nabla_j t_i =0 ,$ then for a
static gravitational field we have $\nabla_i t_j =0$ and hence
$R_{ijk}{^l} t_l = 0, \quad {\tilde R}_{ijk}{^l} = R_{ijk}{^l}. $ With this
it is not difficult to show that a static gravitational field  is absent
($R_{ijk}{^l} =0.$) Thus, the gravitational field  generated by the
gravitational charge  can not be static.

The following point is the  definition of the spherical symmetrical
gravitational potential.  We use the empirical (non internal)
representations about the spherical symmetry. In accordance with this in
the atlas compatible with causal structure a potential with spherical
symmetry has the form $$ dl^2 = A^2 dr^2 + B^2(d{\theta}^2 +\sin^2 \theta
\, d \phi^2), $$ where $A= A(r,\,t), \quad B= B(r,\,t).$ For the nonzero
components of the Ricci tensor  of this one parametric set of metric we
find the following expressions $$ S_{11} = \frac{2}{AB}(A^{'} B^{'} - A
B^{''}), \quad S_{22} = \frac{B}{A^3}(A^{'} B^{'} - A B^{''}) + 1 -
\frac{B^{'2}}{A^2}$$  and $S_{33} = S_{22} \sin^2{\theta},$ where prime
stands for derivative with respect to the variable $r.$ Hence, for
the potential energy of this configuration we have $$ U= \frac{1}{2} S
=\frac{2}{A^3 B}(A^{'} B^{'} - A B^{''}) + \frac{1}{B^2}\left(1 -
\frac{B^{'2}}{A^2}\right). $$  To get $U=0$ we immediately put $   B^{'} =A
$ .  Remind, that we aim  to realize the situation with a pointlike
distribution of the gravitational energy.

For the nontrivial components of the momentum we have
$$P^1_1 = \frac{\dot A}{A}, \quad P^2_2 = P^3_3 = \frac{\dot B}{B}, $$
where dot  denotes partial derivative with respect to the variable
$t$. The flux vector of the energy  has only one nonzero component
$$ G_1 = \frac{2}{AB}(\dot A B^{'} - A {\dot B}^{'})  $$  and in
accordance with relation $A= B^{'}$ the equation $G_i = 0$ is
fulfilled.  Thus,  we have two equations $$
2B { \ddot B} + {\dot B}^2 = 0,\quad { \ddot A} B^2 + {\dot A} {\dot B} B =
\frac{1}{2} A {\dot B}^2.$$ Now we use the other empirical idea about a
spherical wave and put  $$A = A(r-t), \quad B= B(r-t).$$ From the
equation for $B$ only we find that $B {\dot B}^2 = m,$  where $m$ is a
constant of integration ( gravitational charge) and hence $$ B = m
\left[\frac{3}{2} \frac{(r-t)}{m}\right]^{\frac{2}{3}}. $$  Since $A =
B^{'},$ then $$A^{-1} = \left[\frac{3}{2}
\frac{(r-t)}{m}\right]^{\frac{1}{3}} $$ and it is not difficult to see that
second equation for $A$ and $B$ is fulfilled automatically.  Thus, the
Einstein gravitational potential can be presented in the following form
$$d{\tilde s}^2 = - dt^2 + \frac{dr^2} {\left[\frac{3}{2}
\frac{(r-t)}{m}\right]^{\frac{2}{3}} } + \left[\frac{3}{2}
\frac{(r-t)}{m}\right]^{\frac{4}{3}} m^2 (d \theta^2 + \sin^2 \theta d
\phi^2).$$   The solution obtained is known as the Schwarzschild solution
[10].  Since for the kinetic energy we have $T = 0,$ then we see that this
solution really correspond to the situation described above and hence in
definite sense it is an analogous of the Coulomb potential.

Now we consider another very simple but important physical
situation.  The idea about homogenous and isotropic distribution of the
gravitational energy may be realized in the atlas compatible with causal
structure by the one parametric set of metric $$  dl^2 = a(t) d\sigma^2,$$
where $d\sigma^2$ is the metric of the unit 3d sphere, which in the four
dimensional spherical coordinates has the form
$$d\sigma^2 = d\psi^2 + \sin^2\psi(d\theta^2 + \sin^2\theta d\phi^2).$$
Now problem is to find a function $a(t).$  We have
$$S^{\mu}_{\nu} = \frac{2}{a^2} \delta^{\mu}_{\nu}, \quad P^{\mu}_{\nu} =
\frac{\dot a}{a} \delta^{\mu}_{\nu} $$ and hence
for the density of the gravitational energy we get
$$ \varepsilon_g = \frac{3}{a} ({\dot a}^2 +1) .$$  We know that
${\partial}( \sqrt g \,\varepsilon_g) /{\partial t} = 0.  $ Since
$\sqrt g = a^3 \sin^2\psi \sin\theta $ then by integrating over the
variables $\psi, \theta, \phi$ we get the following equation for $a(t)$
from the law of energy conservation $$ a ({\dot a}^2 +1) = 2a_{0} =
const.$$ With respect to the new variable $\eta,$  such that  $dt = a
d\eta$ the solution of the equation in question can be presented in the
following form $a = a_{0}(1- \cos\eta).$ Thus, we get that emergence and
evolution of the clot of gravitational energy is described in  parametric
form as follows $$a = a_{0}(1- \cos\eta), \quad t = a_{0}(\eta- \sin\eta) .
$$ We see that solutions presented here are tightly connected with  the
gravitational energy.   These solutions
have singularities and in connection
with this we would like to make the following remark.

From the theory of time presented here it follows directly
that there is {\bf matter outside the time }. For example, the
gravity field and the electromagnetic field exterior to
time are described by the equations
$$ R_{ij}- \frac{1}{2} g_{ij} R = g_{ij} F^2 - F_{ik} F_{jl} g^{kl},$$
$$ \nabla_i F^{ij} = 0, \quad F^{ij} = F_{kl} g^{ik} g^{jl}, \quad F_{ij}
= \partial_i A_j - \partial_j A_i \  ,$$ where $F^2 = \frac{1}{4} F_{ij}
F^{ij}.$ Recall that the metric $g_{ij}$ is positive definite.
In this context it is very important to understand the nature of
the emergence of time as an objective property of physical systems and as an 
distinctive order parameter.

\section{Conclusion}
Let us now sum up the obtained results  and focus on some
of the problems. It is shown
that gravity physics is an internal field theory by nature which does not
contain apriori elements and can be characterized as follows. In the
theory there is no internal reason that could objectively
distinguish one arbitrary coordinate system from another.
It is unrolled on the smooth four-manifold that does not exist
apriori but is defined by the physical system itself.
Manifold is a basic primary notion in physics and representation
about space and time is given on this ground. All notions, definitions and
laws are formulated in coordinate independent form, i.e., within the
framework of the structure of smooth manifold.
Einstein's gravitational potentials are determined by the positive definite
symmetrical tensor field (Riemann metric) and temporal field.
Being a scalar field on the manifold, the temporal field is
introduced into the theory in the form that provide, in general, time
reversal invariance. The system evolves in its proper time
and it is important that we do not have  compare intrinsic time with
some other times.

 Central role of the time in the internal field
theory is that time determines causal structure of the field theory and
first of all {\bf it characterizes the internal nature of the gravity field as
the simplest closed system}. It is established that energy density is the
first integral of the closed system of interacting fields.
The general covariant definition of the electric and magnetic fields is
given and the Maxwell equations for these fields are derived.
It is shown how to reveal important internal properties
of physical system when coordinates have no physical sense. This means that
physical sense of general covariance is recognized and there exists a
reparametrization-invariant description. And last but not least, it should be
noted that the following problems now become actual: Can gravitational 
charges  be produced on high-energy colliders? Do gravitational charges play
essential role on Earth and in space?
\medskip
%\newpage


\begin{thebibliography}{99} 
\bibitem{1} A. Einstein, M.M. Grossmann, Z. Math. und Phys. {\bf 62},
 225-261 (1913). 
\bibitem{2} A. Einstein,  Ann. Phys. {\bf 49}, 769-822 (1916).
\bibitem{3} C.W.  Misner, K.S. Thorne, and
J.A. Wheeler, {\sl Gravitation}  (Freeman, San Francisco, 1973).
\bibitem{4} A.A.Logunov, {\sl The Theory of Gravity} (Nauka, Moscow, 2001).
\bibitem{5}
B.S. DeWitt, Phys. Rev. {\bf 160}, 1113 (1967).
\bibitem{6}  E. Alvarez,  Rev.
Mod. Phys.  {\bf 61}, 561 (1989).
\bibitem{7}  C. Rovelli, Phys. Rev. D{\bf 43},
442 (1991).\bibitem{8} C. Isham, {\sl Canonical quantum gravity and the problem
of time} (Lectures presented at the NATO Advanced Study Institute,
Salamanca, June 1992); gr-qc/9210011.
\bibitem{9} D. Gromoll, W.Klingenberg, W.Meyer,
 {\sl Riemannsche Geometrie im Grossen} (Springer-Verlag, 1968).
\bibitem{10}  P.A.M.Dirac, {\sl General Theory of Relativity } (A
Wiley-Interscience Publication, 1975), p. 35.
\end{thebibliography}
\end{document}